\documentstyle[12pt]{article}

\def\Journal#1#2#3#4{{#1} {\bf #2}, #3 (#4)}

\def\AN{\em Ann. Phys. (N.Y.)}
\def\ANP{\em Adv. Nucl. Phys.}
\def\ASJ{\em Astr. Jour.}
\def\CPC{\em Comput. Phys. Comun.}

\def\EUA{\em Eur. Phys. J. A}

\def\NPA{{\em Nucl. Phys.} A}
\def\PLB{{\em Phys. Lett.}  B}
\def\PLC{\em Phys. Rep.}

\def\PRL{\em Phys. Rev. Lett.}
 
\def\PRC{{\em Phys. Rev.} C}

\def\RMP{\em Rev. Mod. Phys.}

\def\be{\begin{equation}}
\def\ee{\end{equation}}
\def\bea{\begin{eqnarray}}
\def\eea{\end{eqnarray}}

\begin{document}

\centerline{MICROSCOPIC MODELS for the PARTICLE-VIBRATION COUPLING in EXOTIC NUCLEI}

\centerline{ P.F. BORTIGNON}

\centerline{Institut de Physique Nucl\`eaire, Universit\`e Paris-Sud, F-91406
Orsay Cedex, France,}
\centerline{and Dipartimento di Fisica, Universit\`a di Milano,} 
\centerline{andINFN, Sezione di Milano, Via Celoria 16, I-20133 Milano, Italy}
\centerline{E-mail:
Pierfrancesco.Bortignon@mi.infn.it}

\centerline{abstract}

{Recent results obtained, often in fruitful collaboration with
japanese collegues, in the study of the interplay between single-particle
and collective degrees of freedom in exotic nuclei are reviewed.}

\section{Introduction}

The new renaissance  in nuclear structure claimed by our colleague  I. Tanihata 
in his talk, as produced by the use of radiactive beams and new generations 
of detectors, challenges the theorists with a number of issues. Among these,
the interplay between single-particle and collective degrees of freedom
still needs to be investigated systematically.    

There is no need to emphasize, how strongly the coupling of the 
mean-field single-particle states calculated in the Hartree-Fock (HF)or in the
Hartree-Fock-Bogoliubov (HFB) approach with the  Random Phase Approximation 
(RPA or QRPA) collective surface vibrations 
renormalize the properties of the nuclear excitations, in terms of 
effective masses, charges, spreading widths etc. Decades of works in 
stable nuclei testify it\cite{BM,BBB3,Mah85,Bo98,Paola1}. 

Much less has
been done in nuclei far from the stability line, with some pioneering
works from the groups in Orsay\cite{VM} and Milano\cite{Ghi,CV}.
Recently,
interesting results have been obtained, and they are shortly reported below.

It has been shown\cite{Col0}, that the coupling with density vibrations of the
single-particle HF states in $^{24}$O can lead to new shell structure,
eventually accounting  for the $N$= 16 magic number observed 
experimentally\cite{Oza}.

Giant resonances (GR) are coupled as well to the density vibrations and are known
to acquire a spreading width mainly through this mechanism  in stable nuclei.
In a very recent work\cite{Col1}, the low-lying dipole strength in 
neutron-rich oxygen isotopes $^{18}$O-$^{22}$O
has been calculated and compared to the first experimental data\cite{Aum} 
obtained at GSI using the electromagnetic excitation process at beam
energy around 600 MeV/u on a Pb target. It is concluded that the spreading
is even larger than in stable, heavy nuclei for which a large systematics
exists.

It is well known, that {\it pairing}, the attraction correlating pairs
between the least bound particles in a system, 
 controls almost every aspect of nuclear
stucture close to the ground state. It determines, to a large extent, which nuclei are stable and
which are not.

The very special role played by the pairing force  in drip-line
nuclei, is understood  from the approximate relation between the Fermi level
$\lambda$, pairing gap $\Delta$ and nucleon separation energy $S$
\be
S\approx -\lambda-\Delta.  
\ee
Since for drip-line nuclei $S$ is very small, $\lambda+\Delta \approx 0$,
meaning that the pairing component of the effective interaction can no longer
be treated as a small perturbation important only near the Fermi surface. 
Thus, very recently the simple QRPA calculations performed using zero-range
Skyrme interaction
 in the standard BCS approach\cite{Col2} were compared {\it for the first time} 
 with QRPA results obtained using consistently the finite-range Gogny 
 force\cite{Gog}, not only for the ground state in the HFB approximation,
 but also in solving the QRPA equations to calculate the excited states. 
 
 The richness of the many-body effects in the
 particle-particle channel associated with the coupling to surface vibrations
 is also under rather active study\cite{Jun}.

\section{Results}

\subsection{Giant Resonances}

The calculation in particular for the GR have been done by extending a microscopic
model developed in the last decade within the Milano-Orsay
collaboration\cite{Colo}, in
particular by including the pairing correlations in a simple way.

For this purpose, the effective, energy 
dependent, complex Hamiltonian\cite{Colo}  
\bea
{\cal H}(E)& \equiv & Q_1HQ_1 + W^{\downarrow}(E)\\
& = & Q_1HQ_1 + Q_1HQ_2{1\over {E-Q_2HQ_2+i\epsilon}}Q_2HQ_1,
\eea 
used in closed-shell nuclei, has been generalized and the effects of 
the pairing interaction included, in a Skyrme HF+BCS+QRPA framework\cite{Col2}.
Thus, the eigenstates of $Q_1HQ_1$ are the QRPA eigenstates, including
the low-lying quadrupole and octupole collective vibrations\cite{Col2} to which 
the quasiparticle (qp) states couple in the $Q_2$ space via the $W^{\downarrow}(E)$
term. For each value of the excitation energy $E$,
the QRPA equations for the full ${\cal H}$ are 
solved. The resulting sets of eigenstates $|\nu \rangle$ and complex 
eigenvalues 
$E_\nu-i\Gamma_\nu/2$ enable to calculate all relevant quantities, in  
particular the strength function for an operator $F$
\be
S(E)=-{1\over \pi}\sum_\nu {|\langle \nu | F |g.s.\rangle|^2\over
{E-E_\nu+i\Gamma_\nu/2}}.
\ee
For the dipole strength in the oxygen isotopes $^{18}$O, $^{20}$O
and $^{22}$O, the coupling to the doorway $Q_2$ states (of 4 qp
character), increases  the
low-lying strength below 15 MeV up to 35\%
compared to the QRPA value in the first two isotopes, bringing the cross
section in more than qualitative agreement with
the new data\cite{Aum}, as clearly shown in Table 2 and 3 and Fig. 2 and 3  of the
published work\cite{Col1}. The strong experimental
decrease in $^{22}$O is also reproduced. The same trend is found in a shell 
model calculation\cite{Sag}. Fig. 4  of the same paper\cite{Col1} shows
the strong effect of the
mixing, eventually stronger than in stable, heavier nuclei, because of the
reduced collectivity of the low-lying dipole  strength with few component wave
functions,
reported in Table 3 and 4 of Ref.\cite{Col1} and the asymmetry of the
particle and hole phase-space. All this prevents the strong 
cancellation\cite{BBB3} among the different contributions to the mixing, 
reported for the first time in Fig.1 of Ref. \cite{Col1} for the
quasiparticle case.

It is clear, that a systematic 
appearance and understanding of low-lying dipole strength in neutron-rich nuclei will 
have deep implications in the astrophysical context, e.g. for
the $r$-process, being the statistical (n,$\gamma$) rate related to
the dipole strength function\cite{Gorl}. 

\subsection{Magic Numbers}

The coupling of vibrations to nucleons moving in levels lying close to the
Fermi energy ($E_F$) in atomic nuclei is expected  to lead to a number of 
effects in the s.p. self-energy $\Sigma(E)_j$ added to the mean-field:
(i) shifts of the single-particle levels towards the Fermi energy and thus
an increase of the level density, because of the opposite effects\cite{Mah85}
on the
occupied and unoccupied states of the ''polarization'' and `correlation
contributions\cite{Mah85} (the diagrams in Fig.1 of paper\cite{Col0} 
(ii) single-particle depopulation,
and thus average spectroscopic factors $Z_\omega$ different from unity.

A new frontier of these $\Sigma(E)_j$ calculations is found in the 
physics of the exotic nuclei, facing the experimental evidence for the 
appearance  of new magic numbers, which   starts to be collected.
Theoretically, novel features of $\Sigma_j$ connected to the low-
energy strength of the collective vibrations and to the coupling with 
the continuum were  discussed  in works of the Milano 
group\cite{CV}. The difference with the results in stable nuclei 
may be qualitative, with level inversion and decreasing of the level 
density at $E_F$, again because  of  different cancellation effects 
among the quoted polarization and correlation contributions. 

This is indeed the case, for example, of the nucleus $^{24}$O recently
discussed\cite{Col0}. The s.p. energy of the 1d$_{5/2}$ state is lowered by
the coupling to the collective 2${+}$ states instead of moving up. Moreover, 
the
energy of the 2s$_{1/2}$ state is also lowered by the coupling to the 3${-}$
vibration. These lowering of occupied s.p. states are very specific
because of the blocking of the available phase-space for the other
contributions of Fig.1 in Ref.\cite{Col0}. This results in a small energy gap
between 1d$_{5/2}$ and 2s$_{1/2}$ states and a very large (larger than 6 
MeV) energy gap
of a new magic number $N$=16, as shown in Fig. 3 of Ref.\cite{Col0}.
This will be consistent with the recent
observation of separation energies and interaction cross sections\cite{Oza}.

\subsection{Pairing}

In the revival of nuclear structure studies, produced by the availability of
exotic nuclei, the problem of nuclear pairing has again
become one the
forefront of the theoretical interest. Indeed, the existence of neutron
halo is due to pairing force\cite{GEOR} and in heavier proton rich $N\simeq Z$
nuclei, the proton-neutron  pairing may play an important role.

A key problem in the pairing treatment, is the choice of the force. Since
the gap equation is already a kind  of in-medium two-body Schr\"odinger
equation, one can not use a $G$ matrix which in itself is a solution of the
in-medium two-body problem\cite{Gar}. 
Apart from the use of a simple interaction with
constant matrix elements in a reduced space around the Fermi surface\cite{BM},  
still done e.g. in the above quoted works\cite{Col2,Col1}, it has become
popular in the study of exotic nuclei the use of a density-dependent zero
range- forces with a cutoff, as first introduced in Ref.\cite{GEOR}. It reads
\be
V(\vec r_1,\vec r_2)=V_0(1-x\lbrack\rho((\vec r_1+\vec r_2)/2)/\rho_o
\rbrack^{\alpha})\delta(\vec r_1-\vec r_2),
\ee
depending on 3 parameters ($\rho_0$ is the saturation density)  and
an energy-space cutoff value, otherwise the gap equation
would diverge. Eventually , the cutoff value and $V_0$  may be chosen to
reproduce at zero density the scattering length.

On the other side, the finite-range Gogny force exists\cite{Gogn1,Gogn2},
which allows to performe pairing calculations without introducing  new
parameters and consistenly with the force used in the mean-field.
Although it should be considered as a $G$ matrix, it is found\cite{Gar} that
in the $^1S_0$ channel it acts very much like a realistic bare force,
especially in the D1S version\cite{Gogn2} and at least up to the Fermi energy.

Thus, quite recently\cite{Gog} in our group, we coupled to the HFB
code\cite{Brig} a QRPA code, which allows to calculate for the {\it first
time} the excited states of a superfluid nucleus with the Gogny D1S\cite{Gogn2} interaction.
The continuum is discretized by using a box of appropriate dimensions, (see
the works\cite{Doba,Gras} for a discussion of the role of the continuum in
the pairing problem)  and a
s.p. Wood-Saxon (WS) basis is used in the expansions. All the details will
be found in a large article in preparation. Below,
we will report shortly
the results obtained for the low-lying $2^+$ state and for the dipole
 strength distribution in some O and
Sn isotopes.

In the isotopes $^{18,20,22}$O, the experimental gap values $\Delta(N,Z)$ obtained
with the  formula 
\be
\Delta(N,Z)=1/4(S_n(N+1,Z)+S(N-1,Z)-2S_n(N,Z))
\ee
where $S_n$ is the neutron separation energy, are  nicely reproduced, being
of the order of 2 MeV. However, the results for excitation energies and $B(E2)$
of the first $2^+$, are in strong disagreement with the experimental data,
very much like in the simplified approach HF+Skyrme+BCS of Ref.\cite{Col2}.
The excitation energy are overestimated, being  of the order
of 4 MeV, while the $B(E2$) strongly underestimated. A deeper discussion will
appear in the paper in preparation. Also the dipole strength distribution
is very similar to the one obtained in the work\cite{Col1}, in particular
requiring the coupling to 4qp states to reproduce the experimental strength
below 15 MeV, see the discussion above in the subsection {Giant Resonances}.  

In the Sn isotopes, from A=102 to A=130, the calculated energy of the first $2^+$
is of the order of 2 MeV, the experimental one being 1 MeV, and again the
$B(E2)$ much lower than the known experimental ones. In the 120 and
124 isotopes the centroid enery of the dipole strength over 18 MeV, is about
3 MeV higher than the experimental ones, as already noted in Ref.\cite{Gogn3}
for doubly-magic heavy nuclei, while strength below 15 MeV is clearly found
in $^{132}$Sn, a result of large astrophysical importance\cite{Gorl}.

Having understood the continuum effects\cite{Gras},
much attention is paid
to the many-body effects beyond the mean field (BCS or HFB) approaches, both in
nuclear matter\cite{Bald} and in finite nuclei\cite{Jun}. In the last, in
particular,
are connected, once again, to the coupling of the s.p. motion to the
low-lying surface vibrations (quantal size effects).
While the self-energy effects are expected to reduce the pairing
contributions, because the nucleons spend part of their time in more complicated
configurations ($Z_\omega$ smaller than unity), and to change  level density
(nucleon effective mass $m^*$)\cite{Mah85}, the exchange of collective
surface vibrations between nucleons moving in time reversed states near the
Fermi surface (induced, core polarization
contribution to the effective interaction), is  
found\cite{Barra} to lead to a conspicous contribution to the nuclear
pairing gap. This is very much like the attraction among electrons generated by the
exchange of lattice phonons in the low-temperature
superconductivity\cite{Schr}. 

More quantitatively, we may note that, in the extreme weak coupling limit,
the effective mass $m^*$ and the $Z_\omega$ factor appear in the expression
for $\Delta$ as
\be
\Delta\propto E_F\exp (-{1\over {m^*Z_\omega^2 K^*}})
\ee

A consistent calculation of all many-body  effects, treating self-energy and 
induced pairing interaction on equal footing (not to forget the vertex
corrections), is achieved by solving 
the Dyson equation (also called in this context Nambu-Gor'kov 
equation\cite{Schr}) written as
\be
G^{-1}_j(E)=G^{0-1}_j(E)-\Sigma_j(E).
\ee
Each term is a 2x2 matrix,
and the diagonal and off-diagonal elements of $G^{-1}$ are
the single-particle and pairing (connected to $\Delta$)
Green function respectively. $G^{0-1}$ is the unperturbed one.
In a recent work\cite{Jun},
this Dyson equation
has been solved in the Sn isotopes with phenomenological inputs, a WS with
$m*/m=0.75$, vibrational states reproducing the experimental ones and
standard monopole pairing force with constant matrix element of strength $G$.
It is found\cite{Jun} that the experimental value of the pairing gap of the order of
1.4 MeV is reproduced solving the full Dyson equation with a value of $G$ of
the order of 0.17 MeV against a value of 0.23 MeV in the case of no
particle-vibration coupling.

This approach, applied to the $^{11}$Li 
case\cite{Barra1}, allowed to reproduce the extreme halo properties  of this 
nucleus. In a different context, starting from the results of a shell-model
calculation, the vibrational properties of this nucleus were discussed also in 
Ref.\cite{Suz}.

Any progress done in the description of the pairing correlations in 
extreme conditions (also with respect to the effective interaction to 
be used) will have deep consequences in our 
understanding of the physics of neutron stars, as 
vortices, cooling etc.\cite{NST}.

This work is partly supported by a fellowship from Minist\`ere  de la
Recherche, France. The author wants to thank for the warm hospitality  the
Institut de Physique Nucl\'eaire, where he is spending a  few month stay. 
In
particular he is very grateful to M. Grasso, E. Khan, Nguyen Van Giai, P. Schuck and 
N. Vinh Mau for fruiful discussions.
He likes to thank the many collaborators he had and has on the subjects
discussed, as mentioned in the reference list.





\begin{thebibliography}{99}

\bibitem{BM}
A. Bohr and B.R. Mottelson, {\em Nuclear Structure}, Vol. I and II,
(Benjamin, New York, 1969, 1975).


\bibitem{BBB3} 
G.F. Bertsch, P.F. Bortignon and R.A. Broglia, \Journal{\RMP}{55}{287}{1983}.



\bibitem{Mah85} C. Mahaux, P.F. Bortignon, R.A. Broglia, C.H. Dasso, 
\Journal{\PLC}{120}{1}{1985};
C. Mahaux and R. Sartor,\Journal{\ANP}{20}{1}{1991}. 



\bibitem{Bo98}
P.F. Bortignon, A. Bracco and R.A. Broglia,  {\em Giant Resonances,  Nuclear 
Structure at Finite Temperature}, (Harwood Acad., New York, 1998).


\bibitem{Paola1} 
P. Donati, T. D\o ssing, Y.R. Shimizu, S. Mizutori, P.F.
Bortignon and R.A. Broglia, 
\Journal{\PRL}{84}{4317}{2000}. 

\bibitem{VM} N.Vinh Mau,
\Journal{\NPA}{592}{33}{1995}.

\bibitem{Ghi}
F. Ghielmetti, G. Col\`o, P.F. Bortignon and R.A. Broglia,
\Journal{\PRC}{54}{R2143}{1996}.

\bibitem{CV} G. Col\`o, P.F. Bortignon and R.A. Broglia,
\Journal{\NPA}{649}{335c}{1999}.


\bibitem{Col0}
G. Col\`o, T. Suzuki and H. Sagawa,
\Journal{\NPA}{695}{167}{2001}.

\bibitem{Oza}
A. Ozawa {\it et al.},
\Journal{\PRL}{84}{5493}{2000}. 



\bibitem{Col1}
G. Col\`o and P.F. Bortignon,
\Journal{\NPA}{696}{427}{2001}.


\bibitem{Aum}
A. Leistenschneider {\it et al.},
\Journal{\PRL}{86}{5442}{2001}.


\bibitem{Col2}
E. Khan {et al.},
\Journal{\PLB}{490}{45}{2000}.


\bibitem{Gog}
G. Giambrone, {\em Ph. D.thesis}, (Milano 2001); G. Giambrone {\it et al.}, 
{\em to be published}. 



\bibitem{Jun}
J. Terasaki, F. Barranco, R.A. Broglia, E. Vigezzi and P.F. Bortignon,
\Journal{\NPA}{697}{97}{2001}; arXiv:nucl-th/0109058.


\bibitem{Colo}
G. Col\`o, Nguyen Van Giai, P.F. Bortignon anf R.A. Broglia,
\Journal{\PRC}{50}{1496}{1994}.



\bibitem{Sag}
H. Sagawa and T. Suzuki,  
\Journal{\PRC}{59}{3116}{1999}.
 


\bibitem{Gorl}
S. Goriely,
\Journal{\PLB}{436}{10}{1998}.

\bibitem{GEOR}
G.F. Bertsch and H. Esbensen,
\Journal{\AN}{209}{327}{1991}.

\bibitem{Gar} E. Garrido, P. Sarriguren, E. Moya de Guerra and P. Schuck,
\Journal{\PRC}{60}{064312}{1999}.

\bibitem{Gogn1} 
J. Decharg\'e and D. Gogny,
\Journal{\PRC}{21}{1568}{1980}.

\bibitem{Gogn2}
J.F. Berger, M. Girod and D. Gogny
\Journal{\CPC}{63}{365}{1991}.


\bibitem{Doba}                                            
J. Dobaczewski {\it et al},                                    
\Journal{\PRC}{53}{2809}{1996}.


\bibitem{Gras}
M. Grasso, N. Sandulescu, Nguyen Van Giai, and R.J. Liotta,
arXiv:nucl-th/0109058, \Journal{PRC}{64}{in press}{2001}.



\bibitem{Brig}
S. Briganti,
{\em Thesis} (Milano 1998).

\bibitem{Gogn3}
J. Decharg\'e and L. Sips,
\Journal{\NPA}{407}{1}{1983}.



\bibitem{Bald}                                            
M. Baldo and A. Grasso,                                    
\Journal{\PLB}{485}{115}{2000}.


\bibitem{Barra}                                            
F. Barranco, R.A. Broglia, G. Gori, E. Vigezzi, P.F. Bortignon and J. 
Terasaki,                                     
\Journal{\PRL}{83}{2147}{1999}.

\bibitem{Schr}
J.R. Schrieffer, {\em Theory of Superconductivity}, (Benjamin, New York, 1964).

\bibitem{Barra1}                                            
F. Barranco, P.F. Bortignon, R.A. Broglia, G. Col\`o, E. Vigezzi,                                     
\Journal{\EUA}{11}{385}{2001}.

\bibitem{Suz}
T. Suzuki, H. Sagawa and P.F. Bortignon,
\Journal{\NPA}{662}{282}{2000}.

\bibitem{NST} F. Barranco, R.A. Broglia, H.Espensen and E. Vigezzi,
\Journal{PRC}{58}{1257}{1998};
P.M. Pizzochero,
\Journal{\ASJ}{502}{L153}{1998};
P.M. Pizzochero, F. Barranco, E. Vigezzi and R.A. Broglia, {\em to be
published}.


\end{thebibliography}
\end{document}